\documentclass[pra,aps,reprint,superscriptaddress]{revtex4-1}

\usepackage{amssymb,amsmath}     
\usepackage{graphicx}            
\usepackage{dsfont}				 
\usepackage{xcolor}
\usepackage{qcircuit}

\usepackage{mathtools}
\DeclarePairedDelimiter\bra{\langle}{\rvert}
\DeclarePairedDelimiter\ket{\lvert}{\rangle}
\DeclarePairedDelimiterX\braket[2]{\langle}{\rangle}{#1 \delimsize\vert #2}

\begin{document}

\title{Many-body strategies for multi-qubit gates - \\ quantum control through Krawtchouk chain dynamics}

\author{Koen \surname{Groenland}}
\affiliation{QuSoft, Science Park 123, 1098 XG Amsterdam, the Netherlands}
\affiliation{Inst. of Physics, Univ.~of Amsterdam, Science Park 904, 1098 XH Amsterdam, the Netherlands}
\affiliation{CWI, Science Park 123, 1098 XG Amsterdam, the Netherlands}
\author{Kareljan \surname{Schoutens}}
\affiliation{QuSoft, Science Park 123, 1098 XG Amsterdam, the Netherlands}
\affiliation{Inst. of Physics, Univ.~of Amsterdam, Science Park 904, 1098 XH Amsterdam, the Netherlands}

\date{April 18, 2018}
 
\begin{abstract}
We propose a strategy for engineering multi-qubit quantum gates. As a first step, it employs an {\it eigengate} to map states in the computational basis to eigenstates of a suitable many-body Hamiltonian. The second step employs resonant driving to enforce a transition between a single pair of eigenstates, leaving all others unchanged. The procedure is completed by mapping back to the computational basis.  We demonstrate the strategy for the case of a linear array with an even number $N$ of qubits, with specific $XX+YY$ couplings between nearest neighbors. For this so-called Krawtchouk chain, a 2-body driving term leads to the iSWAP$_N$ gate, which we numerically test for $N=4$ and $6$.
\end{abstract}

\maketitle

\section{Introduction}
The universality of the CNOT plus all 1-qubit gates guarantees that all $N$-qubit unitaries can be composed out of elementary 1-qubit and 2-qubit gates \cite{Nielsen2010}. Nevertheless, the construction of specific multi-qubit gates, such as an $N$-Toffoli gate (a NOT controlled by $N-1$ control qubits) can be cumbersome. As an example, the $N=4$ Toffoli gate employed in a recent implementation of Grover's search algorithm in a trapped ion architecture  \cite{Figgatt2017} employed 11 2-qubit gates derived from the native $XX$ coupling, and 22 1-qubit gates. 

This work proposes an approach towards building $N$-qubit gates which avoids a decomposition into 1-qubit and 2-qubit building blocks. What we propose instead is a protocol which enforces $N$-qubit gates through resonant driving of eigenstates in a suitably engineered quantum many-body spectrum. At first sight such an approach seems hard to achieve. One needs 
\begin{itemize}
\item
an efficient quantum circuit to construct the eigenstates,
\item
a driving term (preferably of 1-qubit or 2-qubit nature) that is resonant with a small number of transitions between eigenstates, and
\item
a way to keep dynamical phases in check, either by tuning the spectrum such that all phases vanish after a known time, or by inverting the spectrum halfway through the protocol. 
\end{itemize}

We here demonstrate how all this can be made to work in the specific setting of a qubit chain with 2-qubit couplings of $XX+YY$ type and adjustable 1-qubit terms. Tuning the couplings to those of the so-called Krawtchouk chain guarantees that the 1-body eigenvalues are all (half-)integer, and the existence of a Jordan-Wigner mapping to non-interacting fermions implies that this property extends to the many-body spectrum. The particular group-theoretic structure of the Krawtchouk operators (which form an \textbf{so}(3) angular momentum algebra) provides the key for the construction of an efficient quantum circuit for a {\it Krawtchouk eigengate} mapping computational states to eigenstates. Finally, the non-local relation between the qubits and the fermion degrees of freedom implies that a driving term involving one or two qubits can connect eigenstates with Hamming distance $N$. By driving resonant to the transition energy, we construct a gate we call iSWAP$_N$, which (for $N$ even) maps states $\ket{1^{N \over 2}0^{N \over 2}}$ and $\ket{0^{N \over 2}1^{N \over 2}}$ onto each other with a phase factor $i$, and acts as identity on all other states. In Appendix \ref{app:circuits} we explain how this gate can be efficiently mapped to more conventional gates, such as a NOT- or iSWAP$_2$-gate with $N-2$ controls.

The remainder of this paper is devoted to this specific example. We stress however, that many variations on the general strategy outlined in the above are possible.

\subsection{Resonant driving}
The prototypical example for resonant driving is a 2-level system with Hamiltonian
\begin{align}
H_D(t) = \left( \begin{array}{cc} E_1 & A \, e^{i\omega t} \\ A \, e^{-i\omega t} & E_2 \end{array} \right) ,
\label{eq:2level}
\end{align}
where we assume that the driving amplitude $A$ is real and positive. We denote by $U_D$ the unitary evolution of quantum states according to Schr\"odinger's equation after a specific time $\tau$. For resonant driving, $\omega=E_2-E_1$, an $H_D(t)$ pulse of duration $\tau_D=\pi / (2A)$ executes the gate $U_D=-iX$ and thus drives the transitions $1 \leftrightarrow 2$ without any error. Off resonance, with $A \ll \Delta=|\omega-(E_2-E_1)|$, the time-evolution stays close to the identity. Putting again $\tau_D=\pi / (2A)$, and assuming that both $\tau_D (E_2-E_1)/(2\pi)$ and $\tau_D \Delta/(2\pi)$ are integers, one finds that the error $\mathcal{E}$ is to leading order given by
\begin{align} 
\mathcal{E} & \equiv 1 - {1 \over 2} | {\rm Tr}\, [ U_D ] | 
\nonumber \\
   & = 1-| \cos [{\pi \over 4A}(\sqrt{\Delta^2 + 4 A^2} )] |
   \nonumber \\
   & \sim {\pi^2 \over 8}\left( {A \over \Delta}\right)^2 = {\pi^4 \over 32 \Delta^2} {1 \over \tau_D^2}.
\end{align}

Below we propose many-body driving protocols, acting on the $2^N$ states of an $N$-qubit register. They have a single resonant transition and stay close to the identity for all other states. We measure the error $\mathcal{E}$ of the driving gate $U_D$ as compared to the target gate $U_{\rm target}$ as

\begin{align}
\mathcal{E} = 1 - {1 \over 2^N} | {\rm Tr} \, [U_{\rm target} U^\dagger_D]| \ .
\label{eq:er}
\end{align}
We will find that this error typically scales as $\tau_D^{-2}$. 

\subsection{$XX+YY$ coupling}
The paper \cite{Schuch2003} analyzed 2-qubit gates based on an $XX+YY$ interaction
\begin{align}
H^{(2)} =- {J \over 4} (X_1 X_2 + Y_1 Y_2).
\label{eq:2qubit}
\end{align}
It observed that the iSWAP$_2$ gate, obtained through a $\tau=\pi/J$ pulse of $H^{(2)}$, is the native gate for this interaction, and that a gate called CNS (CNOT followed by SWAP) can be obtained by combining a single iSWAP$_2$ gate with suitable 1-qubit gates (see Appendix \ref{app:circuits} for details). The paper also proposed a circuit with 10 nearest neighbor iSWAP$_2$ gates realizing a $N=3$ Toffoli gate.

\section{Multi-qubit gates on the Krawtchouk chain}
We now assume a Hamiltonian, acting on $N=n+1$ qubits,
\begin{align}
H = &\sum_{x = 0}^{n-1} \frac{J_{x}(t)}{2} \big( X_{x}  X_{x+1} + Y_{x}  Y_{x+1} \big)  \nonumber \\
&+ \sum_{x=0}^n (\alpha_{x}(t) X_x + \beta_{x}(t) Y_x + \gamma_{x}(t) Z_x) ,
\label{eq:xyham}
\end{align}
where $\{X_x,Y_x,Z_x \}$ denote the Pauli matrices acting on qubit $x$ and $\{ J_x, \alpha_x, \beta_x, \gamma_x \}$ are real, time-dependent functions over which we assume arbitrary and independent control. The specific choice of couplings
\begin{align}
J^K_x = -\frac{J}{2}  \sqrt{ (x+1)(n-x) }
\label{eqn:kccouplings}
\end{align}
gives rise to the so-called Krawtchouk chain Hamiltonian 
\begin{align}
H^K = \sum_{x = 0}^{n-1} {J^K_x \over 2} \big( X_{x}  X_{x+1} + Y_{x}  Y_{x+1} \big) .
\end{align}
The authors of \cite{Christandl2004} observed that applying $H^K$ for a time $\tau = \pi/J$ exactly mirrors the left- and the right sides of the chain, allowing perfect state transfer (PST) between the ends of the chain (see \cite{Bose2008,Nikolopoulos2014} for reviews). Another surprising application is that a $\tau = \pi/J$ pulse, acting on the product state $ |+\rangle^{\otimes N}$, gives the so-called {\it graph state} on a complete graph, which can be turned into a $N$-body GHZ state by 1-qubit rotations (see for example \cite{Clark2005}). For $N$ odd, $N=\pm1 \!\!  \mod \! 4$,
\begin{align}
& |{\rm GHZ}\rangle = \left( \frac{  |0^N \rangle  + |1^N \rangle  }{\sqrt{2}} \right) \nonumber \\
& \quad = e^{\pm i {\pi \over 4} } \exp[-i {\pi \over 4} X]^{\otimes N} \exp [-i {\pi \over J} H^K] |+\rangle^{\otimes N}.
\end{align}
The {\it Krawtchouk eigengates}\ $U_K$ we present below employ a `half-pulse' of duration $\tau=\pi/(2J)$, eq.~(\ref{eq:eigengate}), or rather a pulse combining the Hamiltonian $H^K$ with its dual $H^Z$, eq.~(\ref{eq:eigengatebis}). The half-pulse was previously used in ref. \cite{Alkurtass2014} to generate the specific state $U_K \ket{1010 \ldots10}$ which maximizes block entropy.

\subsection{Analysing the Krawtchouk chain}
The interaction term in the Hamiltonian (\ref{eq:xyham}) conserves the total spin in the $Z$-direction, hence the eigenstates have a well-defined total spin. We may interpret the spin-up excitations as fermionic particles through a Jordan-Wigner (JW) transform \cite{Jordan1928}:
\begin{align}
f^\dagger_x = [ \prod_{j < x} Z_{j} ] \sigma_x^-, \ \ f_x = [ \prod_{j < x} Z_{j} ] \sigma_x^+,
\end{align}
with $\sigma^+_x=(X_x + i Y_x)/2$,  $\sigma^-_x=(X_x - i Y_x)/2$.
Indeed, the operators $f_x$, $f_{x'}^\dagger$ obey canonical {\em anti-commutation} relations. The quadratic terms in (\ref{eq:xyham}) turn into
\begin{align}
H = \sum_{x=0}^{n-1} J_x(t) \left( f^\dagger_x f_{x+1} + \text{h.c.} \right)
\end{align}
and we conclude that the fermions are non-interacting.

Following \cite{Christandl2004} we observe that action of $H^K$ on the Fock space states $\ket{0\dots 010 \dots 0}$ with Hamming weight 1 is the same as the action of the angular momentum operator $L_X$ acting on the spin states of a particle with spin $s={n \over 2}$. Denoting the 1-particle state with the `1' at position $x$ as $\ket{ \{x \}}$, and the spin state with $L_z=m$ as $\ket{m}\rangle$, the identification is
\begin{align}
\ket{\{ x\}} \leftrightarrow \ket{m=x-{n \over 2}}\rangle.
\label{eq:id}
\end{align}
As a consequence, the eigenvalues $\lambda_k$ of 1-particle eigenstates $\ket{\{k \}}_{H^K}$ of $H^K$ make up a linear spectrum 
\begin{align}
\lambda_k = J(k - \frac{n}{2}), \quad k \in \{ 0, \dots, n \}.
\label{eq:spectrum}
\end{align}
The eigenstates $\ket{ \{k \}}_{H^K}$ can be expressed as \cite{Albanese2004}
\begin{align}
\ket{ \{ k \}}_{H^K} = \sum_{x=0}^n \phi^{(n)}_{k,x} \ket{\{ x\}}, \quad
\phi^{(n)}_{k,x} = \sqrt{ \frac{ {{n} \choose {x}}   }{  {{n} \choose {k}} 2^n } } K^{(n)}_{k,x},
\label{eqn:eigenstates}
\end{align}
where $K^{(n)}_{k,x}$ denote Krawtchouk polynomials,
\begin{align}
K^{(n)}_{k,x} = \sum_{j=0}^k (-1)^j {{x}\choose{j}} {{n-x}\choose{k-j}}.
\end{align}
The many-body eigenstates with $q$ particles are created by products of $q$ fermionic modes $c^\dagger_k=\sum_{x=0}^n \phi^{(n)}_{k,x} f^\dagger_x$,
\begin{align}
\ket{ \{ k_1 k_2 \dots k_q \}}_{H^K} = c^\dagger_{k_1} c^\dagger_{k_2} \ldots c^\dagger_{k_q}\ket{0}.
\end{align}
They satisfy
\begin{align}
H^K \ket{ \{ k_1 \dots k_q \}}_{H^K} = \left( \sum_{j=1}^q \lambda_{k_j} \right) \ket{\{ k_1 \dots k_q \}}_{H^K}.
\end{align}
As all eigenvalues are (half-)integer multiples of $J$, all dynamical phases reset after time $\tau = 2 \pi M / J$ for $M$ (even) integer.

\subsection{Quantum circuit for Krawtchouk eigenstates}
We now turn to a construction of an {\it eigengate}: a quantum circuit that efficiently generates the many-body eigenstates from states in the computational basis. Surprisingly, we find two simple circuits that do the job,
\begin{align}
U_K & = \exp \left(- i {\pi \over 2J} H^Z \right)  \exp \left( - i {\pi \over 2J} H^K \right)  \exp \left( - i {\pi \over 2J} H^Z \right)
\label{eq:eigengate} \\
        & = \exp \left(  -i {\pi \over J}\frac{(H^K + H^Z) }{\sqrt{2}} \right).
\label{eq:eigengatebis}
\end{align}
Here $H^Z$ is the operator \cite{Kay2005}
\begin{align}
H^Z = {J \over 2} \sum_{x=0}^n (x-{n \over 2}) (\mathds{1}- Z)_x \ .
\end{align}
Its 1-body spectrum is the same, eq.~(\ref{eq:spectrum}), as that of $H^K$, but the eigenvectors are very different: while $H^Z$ is diagonal on states $\ket{ \{ x_1 x_2  \ldots x_q \} }$ in the computational basis, $H^K$ is diagonal on the Krawtchouk eigenstates $\ket{ \{ k_1 k_2 \ldots k_q \} }_{H^K}$.

The key property is that the operator $U_K$ exchanges the eigenstates of $H^Z$ and $H^K$ and thus performs the change of basis that we are after. Labelling both sets $\{x_1 x_2 \ldots x_q\}$ and $\{k_1 k_2 \ldots k_q\}$ by a binary index $s$ taking values in $\{0,1\}^{n+1}$, we have 
\begin{align}
U_K \ket{s} = i^{q n} \ket{s}_{H^K} \hspace{4mm} \forall s \in \{0,1\}^{n+1}.
\label{eqn:eigengate_summarized}
\end{align}
 
The key property guaranteeing that $U_K$ performs the change of basis is
\begin{align}
H^K U_K =  U_K H^Z .
\label{eq:intertwine}
\end{align}  
This can be established by using that the Krawtchouk operators $H^K$ and $H^Z$ obey \textbf{so}(3) angular momentum commutation relations. We defer the derivation to Appendix \ref{app:grouptheory}, and address the effect of noise in Appendix \ref{app:noisy_eg}. The commutation relations allow us to picture the unitary $U_K$ as a rotation on the Bloch sphere, which agrees perfectly with the Hadamard transformation for $n=1$, $s=10$, $01$, up to a factor $i$.  

\begin{figure}
  \begin{center}
    \includegraphics[width=.48 \textwidth]{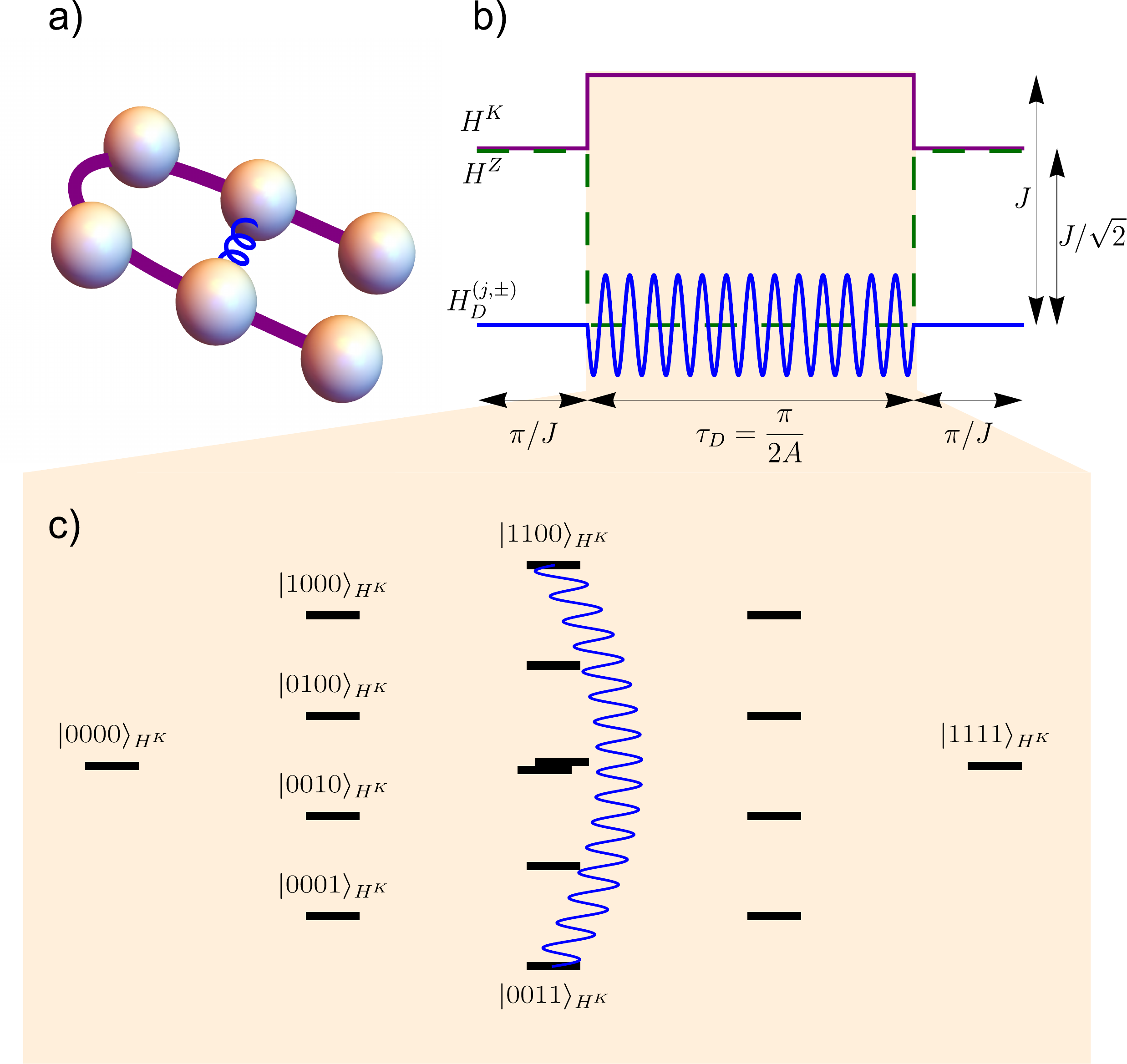}
  \end{center}
  \caption{Protocols for the proposed iSWAP$_N$ gates. a) The $N=6$ qubit chain (spheres) evolving under the Krawtchouk Hamiltonian (solid lines). The driving Hamiltonian $H_D^{(1,-)}$ is depicted as the corkscrew line. b) Field strengths as function of time. c) The spectrum of $H^K$ for $N=4$, with the resonant transition depicted as the curvy line.}
\label{fig:krawtchoukdriving}
\end{figure} 

\subsection{Resonant driving on Krawtchouk eigenstates}
We first assume $N$ odd and consider a driving term coupling $\ket{0^{{n \over 2}+1} 1^{n \over 2}}_{H^K}$ and $\ket{1^{{n \over 2}+1} 0^{n \over 2}}_{H^K}$. The Hamming distance between these two states is $N$. Nevertheless it turns out that the two states can be coupled by a 1-qubit driving term. To see this, we write the JW transform as
\begin{align}
\sigma_x^- = [ \prod_{j < x} (1-2 \hat{n}_{j}) ] f^\dagger_x, \ \  \sigma_x^+ = [\prod_{j < x} (1-2\hat{n}_{j})]f_x
\label{eq:jw}
\end{align}
with $\hat{n}_j = f^\dagger_j f_j$. Targeting the middle qubit, $x={n \over 2}$, we observe that  the operator $\sigma_x^+$ contains precisely the right number of annihilation and creation operators operators to connect the two states. However, we find that amplitude of the matrix element is exceedingly small,
\begin{align}
{}_{H^K} \bra{1^{{n \over 2}+1} 0^{n \over 2}}   \sigma^-_{n \over 2} \ket{0^{{n \over 2}+1} 1^{n \over 2}}_{H^K} = (-2)^{-{n^2 \over 4}}.
\end{align}
Due to this, a resonant driving protocol based on this transition is problematic for $N \geq 5$.

The numbers work out better for a 2-qubit term driving a transition from $\ket{0^{
N \over 2} 1^{N \over 2}}_{H^K}$ to $\ket{1^{N \over 2} 0^{N \over 2}}_{H^K}$ for $N$ even. We propose the driving terms 
\begin{align}
& H^{(j,+)}_D = J_D \cos(\omega t) [\sigma_j^+ \sigma^-_{j+{N \over 2}} + \sigma_j^- \sigma^+_{j+{N \over 2}}],
\nonumber \\
& H^{(j,-)}_D = i J_D \cos(\omega t)  [\sigma_j^+ \sigma^-_{j+{N \over 2}} -  \sigma_j^- \sigma^+_{j+{N \over 2}}].
\end{align}
Note that the locations of the 1-qubit terms are precisely such that, together with the JW string, the required $\frac{N}{2}$ fermion creation \'{a}nd annihilation operators are contained in the driving fields. Making the string any longer would result in effectively less fermionic operators due to symmetry with respect to a global $\prod_x Z_x$ reflection. For $N=6$, we use the `central' 2-qubit driving operator that connects sites $x=1$ and $x=4$, which gives a coupling
\begin{align}
A = \frac{1}{2} | {}_{H^K} \bra{1^3 0^3}  H^{(1,-)}_D(t=0)  \ket{0^3 1^3}_{H^K} | = {5 \over 64} J_D
\label{eq:A}
\end{align}
whereas the largest matrix element of this operator in the 3-particle sector is ${9 \over 32} J_D$. Surprisingly, the matrix elements can be calculated explicitly even for larger $N$, as we show in Appendix \ref{app:mtx_elts}. 

Figure \ref{fig:krawtchoukdriving} depicts the protocol for the resonant driving. Having performed a first Krawtchouk eigengate, taking time $\tau_K=\pi/J$, we turn on the combination
\begin{align}
H^K+H_D^{(1,-)}(t), 
\end{align}
starting at $t=0$, with the driving frequency $\omega=9J$ adjusted to the energy difference between $\ket{0^3 1^3}_{H^K}$ and $\ket{1^3 0^3}_{H^K}$. 
Choosing $\tau_D = 2\pi M/J$ with $M$ integer guarantees that all relative dynamical phases return to unity at time $\tau_D$. Choosing in addition $A={5 \over 64} J_D ={ \pi \over 2 \tau_D}={ J \over 4M }$ leads to a time-evolution that effectuates the transition
\begin{align}
\ket{0^31^3}_{H^K} \rightarrow i \ket{1^30^3}_{H^K}, \ \
\ket{1^30^3}_{H^K} \rightarrow i \ket{0^31^3}_{H^K}.
\end{align}
The protocol is completed by a second Krawtchouk eigengate of time $\tau_K=\pi/J$. Summarizing,
\begin{align}
&  \ket{0^3 1^3} \stackrel{U_K}{\longrightarrow} \ket{0^3 1^3}_{H^K} \stackrel{U_D}{\longrightarrow} i \ket{1^3 0^3}_{H^K} \stackrel{U^\dagger_K}{\longrightarrow} i \ket{1^3 0^3}, 
\nonumber \\
&  \ket{1^3 0^3} \stackrel{U_K}{\longrightarrow} \ket{1^3 0^3}_{H^K} \stackrel{U_D}{\longrightarrow} i \ket{0^3 1^3}_{H^K} \stackrel{U^\dagger_K}{\longrightarrow} i \ket{0^3 1^3}.
\end{align}
A realistic implementation could apply an envelope over all control signals to guarantee smooth evolution of the fields. 

\subsection{The halfway inversion} In numerical simulations, we implemented a spin-echo optimalization, which inverts the many-body spectrum halfway through the driving protocol, such that detrimental dynamical phases accumulated through second-order effects such as Lamb shifts partially cancel. After driving for time $\tau_D/2$, we turn off $H^K$ and turn on $H^Z$ for time  $\pi/J$, which is equivalent to applying a gate of the form $\text{diag}(1,\pm i)$ on each qubit. This effectively performs perfect state transfer on the energy spectrum, mapping indices $k \rightarrow n-k$, or equivalently, a $\pi$-rotation around the $H^Z$-axis of the \textbf{so}(3) Bloch sphere. We complete the driving part of the protocol by driving once more for time $\tau_D/2$ followed by another $H^Z$-pulse of time $\pi/J$. This works without modification if $J \tau_D$ is an integer multiple of $\pi$, and for general $\tau_D$ when the phase of the driving function is adjusted. 

\section{Simulation results} 
\begin{figure}
  \begin{center}
    \includegraphics[width=.45 \textwidth]{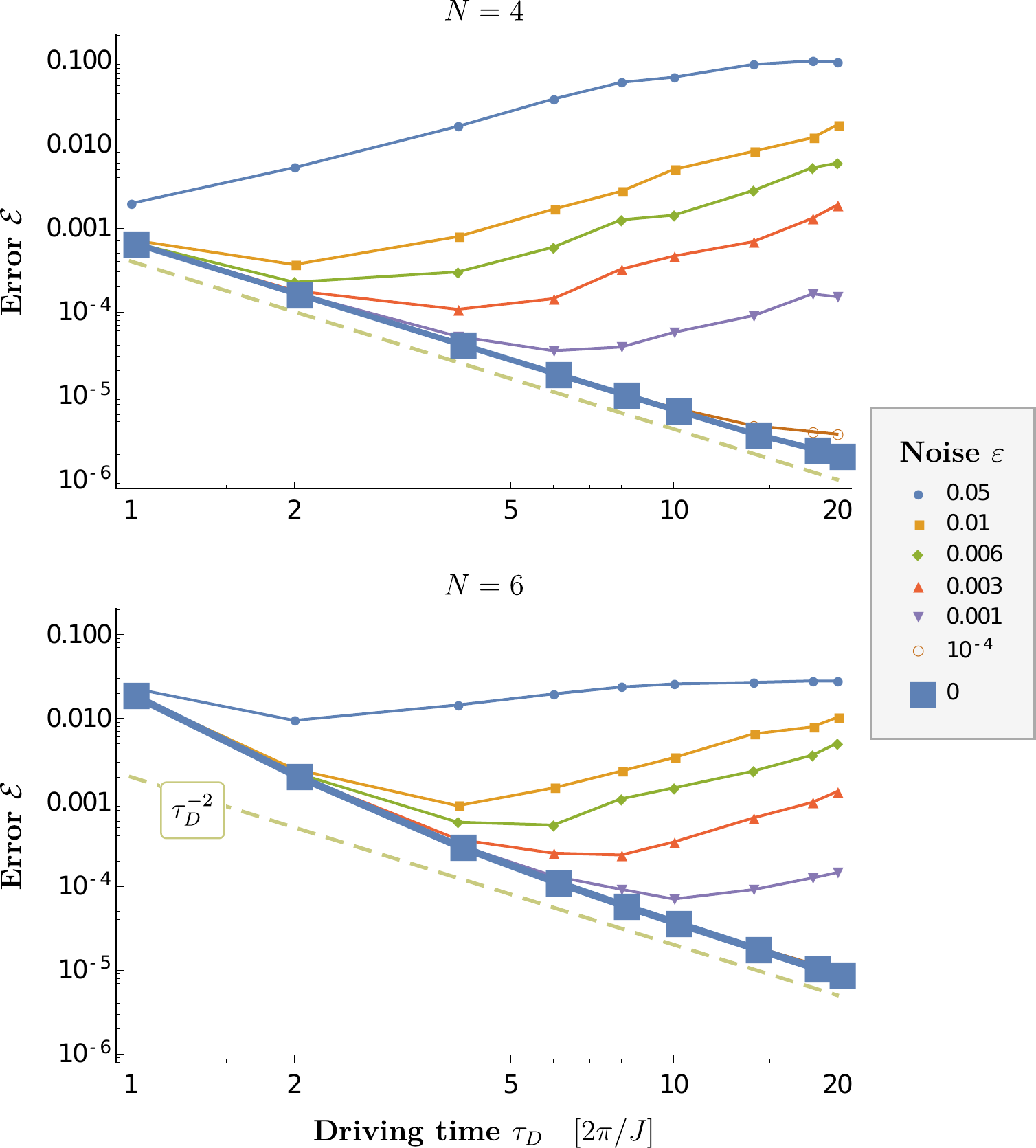}
  \end{center}
  \caption{Fidelities of the resonant driving part of the iSWAP$_4$ and iSWAP$_6$ protocols, including the halfway inversion. The thick lines indicate the errors in the ideal case, thin lines under various values of noise $\varepsilon$. As in the prototypical example eq.~(\ref{eq:2level}), the errors fall off like $\tau_D^{-2}$ (dashed), until the noise $\varepsilon$ becomes the leading source of errors.}
\label{fig:N4N6fidelities}
\end{figure}

Figure \ref{fig:N4N6fidelities} plots the gate-error, defined as in eq.~(\ref{eq:er}), for runtimes up to $M=20$.
The $N=4$ results have been obtained with driving operator 
\begin{align}
H^{(0,+)}_D(t) + H^{(1,+)}_D(t)
\end{align}
with resonant frequency $\omega=4J$. To probe the effect of non-ideal couplings $J_x^K$, we performed the same simulations under multiplicative noise, such that $J_x^K \rightarrow (1+ \varepsilon_x) J_x^K$ where $\varepsilon_x$ is chosen uniformly from $[-\varepsilon, \varepsilon]$. The multiplicative noise is independent of the actual field strengths $J$, making it largely independent of implementation details. The results shown are the averages of at least 180 simulations. 

From figure \ref{fig:N4N6fidelities}, we can read off the time taken by iSWAP$_N$ gates and make a comparison with the time taken by conventional 2-qubit gates derived from the same $XX+YY$-type coupling, see eq.~(\ref{eq:2qubit}). The spatially varying Krawtchouk couplings, eq. \ref{eqn:kccouplings}, grow up to strength $\max_x J^K_x = - \frac{J}{2} \frac{N}{2}$ (for $N$ even), and for a fair comparison we assume the couplings $J^K_x$ may grow no larger than $J^\text{max}/2$ for any $N$. Therefore, we penalize time as a function of $N$ by multiplying by a factor $\frac{N}{2} \frac{J}{J^\text{max}}$. The 2-qubit iSWAP$_2$ gate with coupling maximized at $J^\text{max}/2$ then takes time $\frac{\pi}{J^\text{max}}$. Note that on top of the driving time, the protocol requires 2 eigengates taking unpenalized time $\tau_K = {\pi \over J}$, as well as a halfway inversion consisting of single-qubit gates of the form diag(1,$\pm i$), whose duration we neglect here. We also neglect the error due to a noisy eigengate, which can be seen to be an order of magnitude lower, fig~\ref{fig:eg_error}, than the driving errors encountered here. 

For $N=6$, at sufficiently low noise $\varepsilon < 0.01$, we see an error $\mathcal{E}$ in the order of $10^{-3}$ for $M=4$ meaning it can be achieved in time $\tau=2 \tau_K + \tau_D=10 \pi/J$. Penalizing for the largest couplings being 3 times larger than in the $N=2$ case, we conclude that our iSWAP$_6$ gate takes time equivalent to 30 2-qubit iSWAP$_2$ gates. For $N=4$ an error of well below $10^{-3}$ is already achieved with $M=1$ and we penalize with a factor 2, giving a runtime equivalent of 8 iSWAP$_2$ gates. Note that this is faster than the 10 gates required for 3-qubit Toffoli as proposed in \cite{Schuch2003}.

\section{Implementations}
To our best knowledge, engineered Krawtchouk spin chains have not yet been experimentally tested. Recent experiments \cite{Perez-Leija2013,Chapman2016} report to be the first to engineer Krawtchouk couplings and test PST, but use photonic waveguides which behave different when more than 1 particle is involved. Using NMR, experimental PST was demonstrated on 3 qubits using constant couplings \cite{Zhang2005}, and on up to 6 using iterative procedures \cite{Alvarez2010,Nikolopoulos2014}. However, various theoretical proposals for approximations of Krawtchouk spin chains can be found in literature. The NMR platform could implement spatially varying couplings by using techniques presented in \cite{Ajoy2013}, and numerical tests for this platform have been performed in, for example, ref. \cite{Alkurtass2014}. Alternatively, cold atoms in a 1D optical lattice could be tuned to a regime where a two species Bose-Hubbard description reduces to an $XX+YY$ chain. The authors of \cite{Clark2005} present a numerical study exploring the viability of this scheme to realize graph state generation using Krawtchouk couplings. Another option is to consider superconducting qubits. For those tunable $XX+YY$ couplings are natural, but there is the complication that non-qubit states need to be sufficiently suppressed \cite{Chen2014}.

\section{Conclusion and outlook}
We have outlined a many-body strategy for constructing multi-qubit gates based on driving resonant transitions between many-body eigenstates, and applied it to the example of the Krawtchouk qubit chain. Key in the construction is the {\it eigengate} which maps between the computational basis and the eigenbasis of $H^K$. We applied a simple error model and numerically estimated the fidelity of the protocol. In its current form, our scheme only works for relatively small values of $N$, but we expect optimizations to greatly improve the range of applicability. Moreover, it would be of great interest to find other systems which feature both an eigengate and local driving fields that connect eigenstates, leading to new variations of our protocol.

\section{Acknowledgements}
We thank Anton Akhmerov, Rami Barends, Sougato Bose, Matthias Christandl, Vladimir Gritsev, Tom Koornwinder, Eric Opdam, Luc Vinet, and Ronald de Wolf for enlightening discussions. This research was supported by the QM\&QI grant of the University of Amsterdam, supporting QuSoft.

\onecolumngrid
\appendix

\section{Mapping iSWAP$_N$ to a NOT or iSWAP$_2$ with $N-2$ controls}
\label{app:circuits}
The iSWAP$_N$ gate can be turned into other, more familiar-looking, multi-qubit gates. We first present a circuit which reworks the `double-strength' iSWAP$_N$ gate into an $X$-gate with $N-2$ controls, also known as a generalized Toffoli gate. Doubling the time $\tau_D$ of the resonant driving in our protocol for the iSWAP$_N$ gate leads to a gate that gives minus signs to $\ket{1^{N \over 2}0^{N \over 2}}$ and  $\ket{0^{N \over 2}1^{N \over 2}}$ and leaves all other states put. We combine this gate, which we denote as PHASE$_N$, with an auxiliary qubit initialized to $\ket{0}$, such that only a single state can obtain a sign flip, and finish by conjugating single-qubit gates. For $N=6$, the complete circuit reads
\vskip 3mm
\centerline{
\Qcircuit @C=1 em @R=.6 em {
 & \ctrl{1} & \qw & \raisebox{-6.3 em}{=} & & \qw & \qw & \multigate{5}{{\rm PHASE}_6} & \qw & \qw & \qw \\
 & \ctrl{1} & \qw &                                & & \qw & \qw & \ghost{{\rm PHASE}_6} & \qw & \qw & \qw \\
 & \ctrl{1} & \qw &                                & & \qw & \qw & \ghost{{\rm PHASE}_6} & \qw &  \qw & \qw \\
 & \ctrl{1} & \qw &                                & & \qw & \gate{X} & \ghost{{\rm PHASE}_6} & \gate{X} & \qw & \qw  \\
 & \targ & \qw &                                    & & \gate{H} & \gate{X} & \ghost{{\rm PHASE}_6}  & \gate{X} & \gate{H} & \qw \\
 &               &       &                               & & & \lstick{\ket{0}} & \ghost{{\rm PHASE}_6} & \qw & \lstick{{\ket{0}}} \\
}}
\vskip 6mm
Alternatively, instead of using an ancilla, we may use a modest number of 2-qubit gates in order to form a different $N$-qubit gate. In the main text we mentioned that the native 2-qubit gate for an $XX+YY$ interaction is iSWAP$_2$. Here we show that the multi-qubit gate iSWAP$_N$ can be reworked into an iSWAP$_2$ on the lower two qubits, controled by the other $N-2$ qubits. For concreteness we show the circuit for $N=6$:
\vskip 3mm
\Qcircuit @C=.65 em @R=.2 em {
 & \ctrl{1} & \qw & \raisebox{-5.5em}{=} & & \qw & \qw & \qw & \qw & \multigate{1}{\rm SCN} &\gate{X} & \multigate{5}{{\rm iSWAP}_6} & \gate{X} & \multigate{1}{\rm CNS} & \qw 	& \qw & \qw & \qw & \qw \\
 & \ctrl{1} & \qw & & & \qw & \qw & \qw &\multigate{1}{\rm SCN} & \ghost{\rm SCN} & \qw & \ghost{{\rm iSWAP}_6} & \qw & \ghost{\rm CNS} & \multigate{1}{\rm CNS} & \qw & \qw 	& \qw & \qw \\
 & \ctrl{1} & \qw & & & \qw & \multigate{1}{\rm SCN} & \gate{X} & \ghost{\rm SCN} & \qw & \qw & \ghost{{\rm iSWAP}_6} & \qw & \qw & \ghost{\rm CNS} & \gate{X}	 & \multigate{1}{\rm CNS} & \qw & \qw \\
 & \ctrl{1} & \qw & & & \multigate{1}{\rm SCN} &\ghost{\rm SCN} & \qw & \qw & \qw & \qw & \ghost{{\rm iSWAP}_6} & \qw & \qw & \qw & \qw &\ghost{\rm CNS} & \multigate{1}{\rm CNS} & \qw \\
& \multigate{1}{{\rm iSWAP}_2} & \qw & & & \ghost{\rm SCN} 	&\qw & \qw & \qw & \qw & \qw & \ghost{{\rm iSWAP}_6}  & \qw & \qw & \qw & \qw & \qw & \ghost{\rm CNS}  & \qw \\
& \ghost{{\rm iSWAP}_2} &\qw & & & \qw &\qw & \qw & \qw & \qw & \qw & \ghost{{\rm iSWAP}_6}  & \qw & \qw & \qw & \qw & \qw & \qw & \qw \\
}
\vskip 6mm
In addition to the iSWAP$_6$ gate, the circuit uses 4 CNS gates as well as 4 times the conjugate gate SCN (Swap followed by CNOT). Each of these is obtained from a single iSWAP$_2$ plus 1-qubit gates. For the CNS gate the circuit is \cite{Schuch2003} 
\vskip 3mm
\centerline{
\Qcircuit @C=1em @R=.4em {
& \multigate{1}{\rm CNS} 	& \qw & \raisebox{-2em}{=} 	&& \qw 	& \gate{Z^{-{1 \over 2}}} & \multigate{1}{{\rm iSWAP}_2} 	& \gate{H}  & \qw \\
& \ghost{\rm CNS} 		& \qw & 					&& \gate{H}	& \gate{Z^{-{1 \over 2}}} & \ghost{{\rm iSWAP}_2} 		& \qw  & \qw \\
}}
\vskip 6mm
Following an input state $\ket{111101}$ through the circuit for iSWAP$_2$ with 4 controls, we see that the gates to the left of iSWAP$_6$ send it to $\ket{000111}$. The iSWAP$_6$ gate turns this into $i \ket{111000}$ and the remaining gates produce the output state $i \ket{111110}$. In a similar fashion, $\ket{111110}$ is sent to $i \ket{111101}$. All other states are inert.

For general $N$, the circuit for iSWAP$_2$ with $N-2$ controls uses, in addition to the iSWAP$_N$ gate, $2(N-2)$ iSWAP$_2$ gates plus a number of 1-qubit gates.

\section{Errors due to coupling noise on the eigengate}
\label{app:noisy_eg}
\begin{figure}
  \begin{center}
    \includegraphics[width=.5\columnwidth]{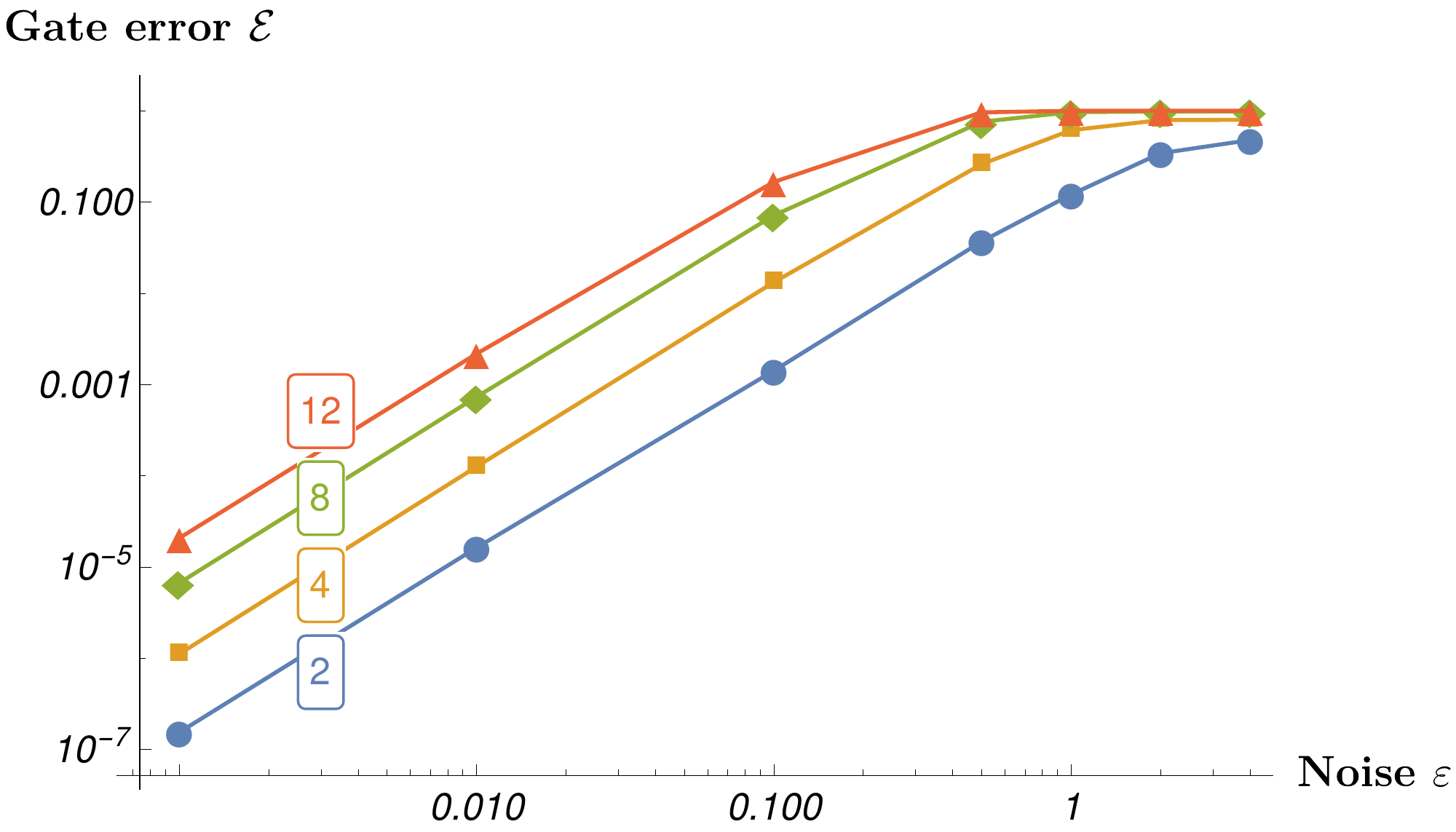}
  \end{center}
  \caption{Numerical results of the trace error compared to the analytical eigengate, for various noise amplitudes, for $N=2,4,8$ and $12$. The error scales roughly as $\mathcal{E} \propto \varepsilon^2$ until the error saturates. Results are averages of at least 110 runs.}
\label{fig:eg_error}
\end{figure}

The eigengate for the Krawtchouk chain is, in principle, analytical and without error. However, it requires couplings $J_x^K$ to be set to an exact number, which is experimentally challenging. Here, we investigate the effect of multiplicative noise on the couplings, such that the actual coupling between qubit $x$ and $x+1$ becomes $J^K_x \rightarrow (1+\varepsilon_x) J_x^K$, with each $\varepsilon_x$ chosen independently and uniformly from $[-\varepsilon, \varepsilon]$. We assume the three-step version of the eigengate is used,
\begin{align*}
U_K & = \exp \left(- i {\pi \over 2J} H^Z \right)  \exp \left( - i {\pi \over 2J} H^K \right)  \exp \left( - i {\pi \over 2J} H^Z \right)
\end{align*}
and that $H^Z$ can be applied without any error. The averages of simulation results, for various $\varepsilon$ and $N$, are displayed in figure \ref{fig:eg_error}. Note that the multiplicative noise is independent of the trade-off between coupling strength $J$ and gate time $\tau = \pi/(2J)$, hence the results are fully general and independent of implementation. Moreover, imprecision in stroboscopic timing is equivalent to some global multiplicative shift in $J_x^K$, hence our results depend strongly on timing errors.
We remark that the errors found above are exceedingly close to the errors of a circuit of depth roughly $N/2+1$ consisting of 2-qubit gates made by the same $XX+YY$ coupling under the same error model. Hence, the eigengate formed by coupling all $N$ qubits at the same time is not any more susceptible to noise than a circuit of moderate depth, and its errors feature the same asymptotic scaling. We aim to make this statement more precise in a future work.

\section{Group theory for the eigengate}
\label{app:grouptheory}
\paragraph{The single $H^K + H^Z$ pulse}
Here, we prove eq. (\ref{eq:intertwine}) of the main text,  
\begin{align*}
H^K U_K =  U_K H^Z,
\end{align*}  
if $U_K$ takes the form 
\begin{align*}
U_K = \exp \left( -i \frac{\pi}{J} \frac{(H^K+H^Z)}{\sqrt{2}}  \right).
\end{align*} 
The identification in eq.~(\ref{eq:id}) inspires the definitions
\begin{align*}
L_X := {1 \over J} H^K, \quad  L_Z := {1 \over J} H^Z,  \quad L_Y := -i [ L_Z, L_X ].
\end{align*}
Indeed, it can be checked that the $L_i$ satisfy the \textbf{so}(3) commutation relations
\begin{align*}
[L_i , L_j] &= i \epsilon_{ijk} L_k, \hspace{1cm} i,j,k \in \{ X,Y,Z \}.
\end{align*}
We now turn to proving 
\begin{align*}
e^{ -i L_H \theta} L_Z e^{ i L_H \theta } = L_X  \quad   \text{ with  }\ L_H :&= \frac{L_X + L_Z}{\sqrt{2}} \nonumber 
\\ \text{ and  }\ \theta &= \pi,
\end{align*}
which explicitly shows that $U_K$ maps eigenstates of $H^K$ to eigenstates of $H^Z$ with corresponding eigenvalues. Because $L_H$ is symmetric in $H^K$ and $H^Z$, the reverse is also true. 

Let $\text{ad}_X Y = [X,Y]$. Then, according to the Baker-Campbell-Hausdorff formula
\begin{align*}
e^A B e^{-A} = e^{\text{ad}_A} B &= B + [A,B] + \frac{1}{2!} [A,[A,B]]  + \frac{1}{3!} [A,[A,[A,B]]] + \dots.
\end{align*}
In our case, we find:
\begin{align*}
e^{ -i L_H \theta} L_Z e^{ i L_H \theta } &= e^{\text{ad}_{-i \theta L_H}} L_Z  \\
&= \sum_{j=0}^\infty \frac{( \text{ad}_{-i \theta L_H} )^j}{j!} L_Z  \\
&= \sum_{j=0}^\infty \frac{(- i \theta )^j}{j!} (\text{ad}_{L_H} )^j L_Z.
\end{align*}
For $1\leq j\leq3$, we calculate the commutators as follows:
\begin{align*}
[L_H,L_Z] &= \frac{-i}{\sqrt{2}} L_Y && (j=1) \nonumber \\
[L_H,[L_H,L_Z]] &= \frac{-i}{2} \left( [L_X,L_Y] + [L_Z,L_Y] \right) = \frac{1}{2} (L_Z-L_X) && (j=2) \nonumber \\
(\text{ad}_{L_H})^3 L_Z &= \frac{1}{2 \sqrt{2}} \left( [L_X+L_Z, L_Z-L_X] \right) =  \frac{-i}{\sqrt{2}}L_Y  &&(j=3) .
\end{align*}
Note that subsequent application of $\text{ad}_{L_H}$ on $L_Z$ causes oscillations between two distinct results. Separating odd and even $j$, and treating $j=0$ as a special case, we find
\begin{align*}
e^{ -i L_H \theta} L_Z e^{ i L_H \theta } &= \sum_{j=0}^\infty \frac{(-1)^j \theta ^{2j}}{(2j)!} (\text{ad}_{L_H} )^{2j} L_Z + \sum_{j=0}^\infty \frac{(-i) (-1)^j  \theta ^{2j+1}}{(2j+1)!}  (\text{ad}_{L_H} )^{2j+1} L_Z 
\nonumber \\
	&= \sum_{j=0}^\infty \frac{(-1)^j \theta ^{2j}}{(2j)!} \left( \frac{L_Z-L_X}{2} \right) + \underbrace{ \left( \frac{L_Z+L_X}{2} \right) }_{\text{compensates } j=0 \text{ term} }  +  \sum_{j=0}^\infty (-i) \frac{(-1)^j \theta ^{2j+1}}{(2j+1)!} \left( \frac{-i L_Y}{\sqrt{2}} \right)  
\nonumber \\
	&= \frac{L_Z+L_X}{2} + \frac{\cos(\theta)}{2} ( L_Z-L_X )  - \frac{\sin(\theta)}{\sqrt{2}} L_Y 
\nonumber \\
	&= \sin^2(\theta/2) L_X - \frac{\sin(\theta)}{\sqrt{2}} L_Y + \cos^2(\theta/2) L_Z.
\end{align*}
Eq.~(\ref{eq:intertwine}) of the main text is recovered when $\theta = \pi$. 

The presented derivation holds even when the  \textbf{so}($3$) commutation relations are replaced by the more general requirement $(\text{ad}_{L_H})^2 (L_Z-L_X) = L_Z-L_X$. This opens up the question which other systems feature an eigengate through continuous evolution. 

\paragraph{The three-step pulse}
To show that $U_K$ functions as an eigengate for the three-pulse variant,
\begin{align*}
U_K = \exp \left(- i {\pi \over 2J} H^Z \right)  \exp \left( - i {\pi \over 2J} H^K \right)  \exp \left( - i {\pi \over 2J} H^Z \right),
\end{align*}
one could employ the same strategy as used in the previous section. However, here we present an alternative perspective, which connects to the theory of orthogonal polynomials. The actions of the exponentials on eigenstates with a single excitation at location $x$ or $k$ is 
\begin{align*}
\exp \left(- i {\pi \over 2J} H^Z \right) \ket{ \{ x \} } = (-i)^{x-n/2} \ket{\{x\}}, \\
\exp \left(- i {\pi \over 2J} H^K \right) \ket{\{k\}}_{H^K} = (-i)^{k-n/2} \ket{\{k\}}_{H^K}.
\end{align*}
Together with the known single-excitation basis transform, eq. (\ref{eqn:eigenstates}), we rewrite the action of $U_K$ as
\begin{align*}
U_K \ket{\{x\}} &= \sum_{k, y} (-i)^{x + y + k - 3n/2} \phi^{(n)}_{x, k} \phi^{(n)}_{k, y} \ket{\{y\}}.
\end{align*}
To prove that this is indeed equal to $i^n \ket{x}_{H^K} = i^n \sum_{y} \phi^{(n)}_{x, y} \ket{y}$, we require the identity
\begin{align*}
(-i)^{x + y - n/2} \sum_{k=0}^n  (-i)^k \phi^{(n)}_{x,k} \phi^{(n)}_{k,y} = \phi^{(n)}_{x, y},
\end{align*}
or equivalently, 
\begin{align*}
\sum_{k=0}^n  (-i)^k   K^{(n)}_{x,k} K^{(n)}_{k,y} &=  (i)^{x + y - n/2} \ 2^{(n/2)} \    K^{(n)}_{x, y}.   
\end{align*}
The latter formula is a special case of Meixner's expansion formula (see equation (3.5) in \cite{Rosengren1999}) after substituting $z \rightarrow i; \   x, y \rightarrow 2; \ \alpha, \beta, \nu \rightarrow -x, -y, -n$. 

For states with more than one excitation we argue that, since particles are non-interacting throughout each of the three pulses, we may apply the above reasoning for each particle independently. We conclude that $U_K \ket{s} \propto \ket{s}_{H^K}$ for all $s \in \{0,1\}^{n+1}$.

\section{Matrix elements of driving operators}
\label{app:mtx_elts}

Here, we derive a more explicit form of the two matrix elements
\begin{align*} 
& M^{(1)}_j = _{H^K}\bra{ 1^{{n\over 2}+1} \ 0^{n \over 2} \ } \sigma_j^- \ket{\ 0^{ {n\over 2}+1} \  1^{n\over 2}}_{H^K} && (N \text{ odd}), 
\\[2mm]
& M^{(2)}_{j,d} = _{H^K}\bra{1^{N\over 2} \ 0^{N\over 2} \ } \sigma_j^- \sigma_{j+d}^+ \ket{ \ 0^{N\over 2} \  1^{N\over 2}}_{H^K} && (N \text{ even}),
\end{align*}
which determine the duration of the resonant transitions described in the main text. The matrix elements can be calculated exactly by rewriting the expressions in terms of fermionic operators. Using the eigenbasis-operators (eq. \ref{eqn:eigenstates}) and keeping only the terms that create \emph{and} annihilate the required number of particles, one obtains
\begin{align*}
M^{(1)}_{j={n \over 2}} = 2^{n \over 2}  \begin{vmatrix} \phi^{(n)}_{ \{0,\ldots, {n \over 2} \} , \{ 0, \ldots, {n \over 2} \} } \end{vmatrix}  \begin{vmatrix}  \phi^{(n)}_{ \{0,\ldots,{n \over 2}-1\},\{{n \over 2}+1,\ldots,n\} } \end{vmatrix},
\end{align*}
where $\begin{vmatrix}  \phi_{\vec{x},\vec{y}} \end{vmatrix}$ denotes the minor of matrix $\phi$ with only rows $\vec{x}$ and columns $\vec{y}$ kept. 
Using
\begin{align*}
\begin{vmatrix} K^{(n)}_{ \{0,\ldots, {n \over 2} \} , \{ 0, \ldots, {n \over 2} \} } \end{vmatrix} &= (-2)^{n(n+2) \over 2} \\
\begin{vmatrix} K^{(n)}_{ \{0,\ldots,{n \over 2}-1\},\{{n \over 2}+1,\ldots,n\} } \end{vmatrix} &= (-2)^{n(n-2) \over 2}
\end{align*}
we find
\begin{align*}
M^{(1)}_{j={n \over 2}} = (-2)^{-n^2/4}.
\end{align*}
Similarly we find
\begin{align*}
M^{(2)}_{j,d={n+1 \over 2}} = & \, 2^{n-1 \over 2} \begin{vmatrix} \phi^{(n)}_{ \{ j, \ldots, j+{n-1 \over 2}\} ,\{0,\ldots,{n-1 \over 2}\} } \end{vmatrix}  \begin{vmatrix}  \phi^{(n)}_{ \{j+1,\ldots,j+{n+1 \over 2} \},\{ {n+1 \over 2},\ldots, n\}} \end{vmatrix},
\end{align*}
which, together with
\begin{align*}
\begin{vmatrix}   K^{(n)}_{ \{ j, \ldots ,j+d-1\} , \{0, \ldots, \frac{n-1}{2} \} } \end{vmatrix} &= (-2)^{(n-1)(n+1) \over 8}, \\
\begin{vmatrix}  K^{(n)}_{ \{ j+1, \ldots,j+d\} , \{\frac{n+1}{2}, \ldots, n \} } \end{vmatrix} &= (-1)^{{(j+1)(n+1) \over 2}}(-2)^{(n-1)(n+1) \over 8} ,
\end{align*}
leads to closed-form expressions for the matrix elements $M^{(2)}_{j,d={n+1 \over 2}}$. For $n=5$ one finds $M^{(2)}_{j=1,d=3}=5/64$ while for $n=3$ we have $M^{(2)}_{0,2}=M^{(2)}_{1,2}=\sqrt{3}/8$.

We stress that for $n$ large both $M^{(1)}$ and $M^{(2)}$ fall off rapidly with $n$. For example, for $N=2,6,\ldots$, putting $j={n-1 \over 4}$, we find asymptotic behavior
\begin{align*}
M^{(2)}_{j={n-1 \over 2},d={n+1 \over 2}} \sim c_0 \,  c_1^n \, c_2^{n^2} n^{-1/6}
\end{align*}
with $c_2=2^{3/4}3^{-9/16}= 0.9065\ldots$. This implies that the run-time of the resonant driving protocol (in its current form) increases rapidly with $n$.

Lastly, we note that matrix elements of the conjugates of the discussed driving fields may have a different phase. In particular, for $N$ even, we find 
\begin{align*}
_{H^K}\bra{1^{N\over 2} \ 0^{N\over 2}  \ } \sigma_j^+ \sigma_{j+{N \over 2}}^- \ket{ \ 0^{N\over 2} \ 1^{N\over 2}}_{H^K}  = (-1)^{N\over 2} M^{(2)}_{j,{N \over 2}}.
\end{align*}
Hence, to achieve constructive interference, the optimal driving terms are of the form $H_D^{(j,+)}$ if $N/2$ is even, or $H_D^{(j,-)}$ if $N/2$ is odd.

\end{document}